# Analysis of half-spin particle motion in Kerr-Newman field by means of effective potentials in second-order equations

V.P.Neznamov[*], V.E.Shemarulin

FSUE "RFNC-VNIIEF", Russia, Sarov, Mira pr., 37, 607188

Abstract

The self-conjugate Dirac Hamiltonian is obtained in the Kerr-Newman field. A transition is implemented to a Schrödinger-type relativistic equation. For the case where the angular and radial variables are not separated, the method of obtaining effective potentials is generalized. Effective potentials have isolated singularities on the event horizons as well as at certain parameters of the Kerr-Newman field and of the fermion in the neighborhoods of some values of the radial coordinate. For the extreme Kerr-Newman field, the impossibility of existence of stationary bound states of half-spin particles is proved.

---

[*] E-mail: vpneznamov@vniief.ru; vpneznamov@mail.ru;

# 1. Introduction

Currently, there are a lot of studies of properties and solutions to the Dirac equation in curved space-time (see, e.g., [1] - [25]). The papers [6], [9] - [18], [26] - [28] are devoted to studies of the Dirac equation in Kerr-Newman space-time. In [29], [30] , the Dirac Hamiltonian was studied in the Kerr-Newman field with "zero" gravity.

Quantum mechanics of half-spin particle motion in external fields can be analyzed using the relativistic Schrödinger-type equation with effective potentials. In this case, upon separation of variables, the system of first-order ordinary differential equations for radial wave functions is transformed into Schrödinger-type equations with effective potentials. Each of these equations is related to only one of two radial wave functions. When analyzing a Schrödinger-type equation, one can use broad experience in studying such equations in nonrelativistic quantum mechanics.

In [31] - [33], the method of effective potentials was used to analyze the motion of electrons and positrons in the Coulomb field. In [34] - [37], this method was applied to analyze the motion of Dirac particles in the Schwarzschild and Reissner-Nordström gravitational fields and in the field of naked singularities of the static $q$-metric [38].

In the present paper, the method of effective potentials is used to analyze singularities of half-spin particle motion in the charged axially symmetric Kerr-Newman field. We examine the Kerr-Newman fields with two horizons, the extreme Kerr-Newman fields with a single horizon and Kerr-Newman naked singularities. For such analysis, the self-conjugate Dirac Hamiltonian is obtained, and, as in [37], the method of obtaining effective potentials in the Schrödinger-type equation is generalized to the case where radial and angular variables are not separated.

As a result, we obtain that the effective potentials have isolated singularities on the horizons and at certain parameters of the Kerr-Newman field and Dirac particle at some intermediate points along the radius.

An important factor is the absence of any features in the self-conjugate Hamiltonian, radial equations and effective potential associated with presence of the ergosphere in whose volume the component of the metric tensor $g_{00}$ is smaller or equal to zero.

The paper is organized as follows. In Section 2, the basic properties of the Kerr-Newman metric are presented. In Section 3, the self-conjugate Hamiltonian in the Kerr-Newman field with a "flat" scalar product of wave functions is generalized. In Section 4, the method of obtaining effective potentials in a Schrödinger-type equation is generalized. In Section 5, singularities of the effective potentials are studied. In the Conclusions, the obtained results are briefly discussed.

Below, as a rule, we use the system of units $\hbar = c = 1$; the signature of the Minkowski space is selected to be

$$\eta_{\underline{\alpha}\underline{\beta}} = diag[1,-1,-1,-1]. \tag{1}$$

In (1) and below, the underlined indices are local ones.

## 2. Kerr-Newman metric

The Kerr-Newman solution is characterized by a point source with mass $M$ and charge $Q$ rotating with the angular momentum $\mathbf{J} = Mc\mathbf{a}$, where $c$ is the velocity of light. The Kerr-Newman metric in Boyer-Lindquist coordinates $(t, r, \theta, \varphi)$ can be presented as

$$ds^2 = \left(1 - \frac{r_0 r - r_Q^2}{r_K^2}\right)dt^2 + \frac{2a(r_0 r - r_Q^2)}{r_K^2}\sin^2\theta dt d\varphi - \frac{r_K^2}{\Delta_{KN}}dr^2 - r_K^2 d\theta^2 - \left(r^2 + a^2 + \frac{a^2(r_0 r - r_Q^2)}{r_K^2}\sin^2\theta\right)\sin^2\theta d\varphi^2. \tag{2}$$

In (2), $r_0 = 2GM/c^2$ is the gravitational radius (event horizon) of the Schwarzschild field; $G$ is the gravitational constant; $r_Q = \sqrt{G}Q/c^2$; $r_K^2 = r^2 + a^2\cos^2\theta$;

$$\Delta_{KN} = r^2 f_{KN} = r^2\left(1 - \frac{r_0}{r} + \frac{r_Q^2}{r^2} + \frac{a^2}{r^2}\right).$$

The equality $g_{00} = 0$ determines the inner and outer surfaces of the ergosphere of the Kerr-Newman field:

$$g_{00} = 1 - \frac{r_0 r - r_Q^2}{r_K^2} = 0. \tag{3}$$

In the ergosphere volume, bounded by the surfaces (3),

$$g_{00} \leq 0. \tag{4}$$

At $Q = 0$, $(r_Q = 0)$, the Kerr-Newman solution turns into the Kerr solution.

If $r_0 > 2\sqrt{a^2 + r_Q^2}$, then

$$f_{KN} = \left(1 - \frac{r_+}{r}\right)\left(1 - \frac{r_-}{r}\right), \tag{5}$$

where $r_\pm$ are radii of the inner and outer event horizons of the Kerr-Newman field,

$$r_\pm = \frac{r_0}{2} \pm \sqrt{\frac{r_0^2}{4} - a^2 - r_Q^2}. \tag{6}$$

The case of $r_0 = 2\sqrt{a^2 + r_Q^2}$, $r_+ = r_- = r_0/2$ corresponds to the Kerr-Newman extreme field.

The case of $r_0 < 2\sqrt{a^2 + r_Q^2}$ corresponds to a naked singularity of the Kerr-Newman field. In this case, $f_{KN} > 0$, and $r \in (0, \infty)$ is the domain of wave functions.

Below, we will analyze the behavior of the effective potentials in the Schrödinger-type equation in the Kerr-Newman field. The Schrödinger-type equation with a self-conjugate Hamiltonian is obtained by squaring the Dirac equation written in the Hamiltonian form. The initial Dirac Hamiltonian must also be self-conjugate.

## 3. Self-conjugate Hamiltonian of a half-spin particle in the Kerr-Newman field

The desired Hamiltonian can be determined using algorithms for obtaining self-conjugate Dirac Hamiltonians in external gravitational fields [21] - [23] by methods of pseudo-Hermitian quantum mechanics [39] - [41].

In [21] - [23], it has been proved that all stationary Dirac Hamiltonians in curved space-time are pseudo-Hermitian. The condition of pseudo-hermiticity of the Hamiltonians $H$ assumes the existence of an invertible operator $\rho$ satisfying the relation

$$\rho H \rho^{-1} = H^+. \qquad (7)$$

The plus superscript in (7) means the Hermitian conjugate. If there exists an operator $\eta$ satisfying the relation

$$\rho = \eta^+ \eta, \qquad (8)$$

then for time-independent Hamiltonians we get a Hamiltonian in the $\eta$-representation

$$H_\eta = \eta H \eta^{-1} = H_\eta^+, \qquad (9)$$

which is self-conjugate, with the spectrum of eigenvalues coinciding with that of the initial Hamiltonian $H$.

The scalar product of the wave functions in the initial representation is

$$\langle \varphi, \psi \rangle_\rho = \int d\mathbf{x} \left( \varphi^+ \rho \psi \right). \qquad (10)$$

The scalar product in the $\eta$-representation has the form standard for quantum mechanics (the flat scalar product)

$$(\Phi, \Psi) = \int d^3x \left( \Phi^+ \Psi \right). \qquad (11)$$

In (10), the operator $\rho$ coincides with the Parker weight operator [43] in the scalar product of the wave functions.

It is evident that with (8) and the transformation $\Psi = \eta\psi$, the scalar products of (10), (11) are

$$\langle \varphi, \psi \rangle_\rho = (\Phi, \Psi). \tag{12}$$

In quantum mechanics, for the existence of the eigenfunctions and the eigenvalue spectrum of the Hamiltonian $H$, it is necessary that holds the Hermiticity condition

$$(\Phi, H\Psi) = (H\Phi, \Psi). \tag{13}$$

For the fulfillment of (13), in addition to self-adjointless of the Hamiltonian, necessary is the appropriate behaviour of the wave functions in the scalar products.

The choice of different systems of tetrad vectors for some physical system can lead to different forms of self-conjugate Hamiltonians in the $\eta$-representation. However, all of them are connected by unitary transformations due to spatial rotations of the Dirac matrices. It is evident that such Hamiltonians are physically equivalent [42]. In [21] - [23], the self-conjugate Dirac Hamiltonians with flat scalar products have been obtained for different spherically symmetric metrics. In [23], a self-conjugate Hamiltonian was also obtained for the axially symmetric Kerr metric with $Q = 0$. In [24], for the Kerr metric, the equivalence of the Hamiltonian from [23] in the representation with a flat scalar product of wave functions was proved with the Chandrasekhar Hamiltonian [4], [5] in the representation with the Parker factor [43] in the scalar product of wave functions.

The Hamiltonian for the Kerr-Newman metric is easily obtained from the Hamiltonian expression for the Kerr metric [23] by substituting of $g^{00} \to g_{KN}^{00}$ and $\Delta_K \to \Delta_{KN}$, where

$$g_{KN}^{00} = \frac{1}{\Delta_{KN}}\left(r^2 + a^2 + \frac{a^2(r_0 r - r_Q^2)}{r_K^2}\sin^2\theta\right), \tag{14}$$

$$\Delta_{KN} = r^2 f_{KN}\left(1 - \frac{r_0}{r} + \frac{a^2 + r_Q^2}{r^2}\right). \tag{15}$$

For charged particles, the Hamiltonian should be completed with a term due to Coulomb interaction between a particle with charge $q$ and the source of the Kerr-Newman electric field.

As a result, the self-conjugate Hamiltonian $H_\eta = H_\eta^+$ for the Kerr-Newman field in the representation with a flat scalar product of wave functions has the form

$$H_\eta = \frac{m}{\sqrt{g_{KN}^{00}}}\gamma^0 - \frac{i\sqrt{\Delta_{KN}}}{r_K\sqrt{g_{KN}^{00}}}\gamma^0\gamma^1\left(\frac{\partial}{\partial r}+\frac{1}{r}\right) - \frac{i}{r_K\sqrt{g_{KN}^{00}}}\gamma^0\gamma^2\left(\frac{\partial}{\partial\theta}+\frac{1}{2}\cot\theta\right)-$$
$$-\frac{i}{g_{KN}^{00}\sqrt{\Delta_{KN}}\sin\theta}\gamma^0\gamma^3\frac{\partial}{\partial\varphi} - \frac{i}{g_{KN}^{00}}\frac{ar_0 r}{r_K^2\Delta_{KN}}\frac{\partial}{\partial\varphi} - \frac{i}{2}\gamma^0\gamma^1\frac{\partial}{\partial r}\left(\frac{\sqrt{\Delta_{KN}}}{r_K\sqrt{g_{KN}^{00}}}\right)-$$
$$-\frac{i}{2}\gamma^0\gamma^2\frac{\partial}{\partial\theta}\left(\frac{1}{r_K\sqrt{g_{KN}^{00}}}\right) + \frac{i}{4}\gamma^3\gamma^1\sqrt{g_{KN}^{00}}\frac{\Delta_{KN}}{r_K}ar_0\sin\theta\frac{\partial}{\partial r}\left(\frac{r}{g_{KN}^{00}r_K^2\Delta_{KN}}\right)-$$
$$-\frac{i}{4}\gamma^2\gamma^3\sqrt{g_{KN}^{00}}\frac{\sqrt{\Delta_{KN}}}{r_K}ar_0\sin\theta\frac{\partial}{\partial\theta}\left(\frac{r}{g_{KN}^{00}r_K^2\Delta_{KN}}\right) + \frac{qQ}{r},$$ (16)

where $\gamma^\mu$ are Dirac matrices with local indices satisfying standard anticommutation relations

$$\gamma^\mu\gamma^\nu + \gamma^\nu\gamma^\mu = \eta^{\mu\nu}.$$

The Dirac equation with the Hamiltonian (16) has the form

$$i\frac{\partial\Psi(\mathbf{r},t)}{\partial t} = H_\eta\Psi(\mathbf{r},t).$$ (17)

## 4. Effective potentials for the Kerr-Newman field

It can be seen from the form of the Hamiltonian (16) that the radial and angular variables $(r,\theta)$ in Eq. (17) are not separated. It is necessary to generalize the standard method of obtaining effective potentials by squaring Dirac equations.

Let us represent the wave function $\Psi(\mathbf{r},t)$ in (17) as

$$\Psi(\mathbf{r},t) = \begin{pmatrix} \omega(r,\theta)\xi(\theta) \\ -i\chi(r,\theta)\sigma^3\xi(\theta) \end{pmatrix} e^{-iEt}e^{im_\varphi\varphi},$$ (18)

where $E$ is the energy of the Dirac particle, $m_\varphi$ is the magnetic quantum number, and the spinor $\xi(\theta) = \begin{pmatrix} _{-1/2}Y(\theta) \\ _{+1/2}Y(\theta) \end{pmatrix} \equiv \begin{pmatrix} Y_-(\theta) \\ Y_+(\theta) \end{pmatrix}$ represents spherical harmonics for a half spin. The explicit form $\xi(\theta)$ can be presented as in [44]:

$$\xi(\theta) = \begin{pmatrix} Y_{-jm_\varphi}(\theta) \\ Y_{+jm_\varphi}(\theta) \end{pmatrix} = (-1)^{m_\varphi+1/2}\sqrt{\frac{1}{4\pi}\frac{(j-m_\varphi)!}{(j+m_\varphi)!}}\begin{pmatrix} \cos\theta/2 & \sin\theta/2 \\ -\sin\theta/2 & \cos\theta/2 \end{pmatrix}\times$$
$$\times\begin{pmatrix} (\kappa-m_\varphi+1/2)\ P_l^{m_\varphi-1/2}(\theta) \\ P_l^{m_\varphi+1/2}(\theta) \end{pmatrix},$$ (19)

where the expression after the square root in parentheses is a two-dimensional matrix;

$P_l^{m_\varphi \pm 1/2}$ are Legendre associated functions; $j, l$ are the quantum numbers of total angular and orbital momenta of the Dirac particle, $m_\varphi = -j, -j+1, ..., j$.

Besides, let us note the following two facts:

1. Since the variables $(r, \theta)$ in (17) are not separated, the functions $\omega(r, \theta)$ and $\chi(r, \theta)$ depend on $r$ and $\theta$.

2. To obtain real effective potentials, it is necessary for the functions $\omega(r, \theta)$ and $\chi(r, \theta)$ to be real as well.

Upon substitution of (18), Eq. (17) will contain the spinors $\xi(\theta)$ and $d\xi(\theta)/d\theta$, the functions $\omega(r, \theta)$ and $\chi(r, \theta)$, and their derivatives with respect to $r$ and $\theta$.

If an equivalent substitution is implemented in the Hamiltonian (16)

$$\gamma^1 \to \gamma^3, \; \gamma^3 \to \gamma^2, \; \gamma^2 \to \gamma^1, \tag{20}$$

the derivative $d\xi(\theta)/d\theta$ in (17) can be removed using the Brill-Wheeler equation [45]

$$\left[ i\sigma^2 \left( \frac{\partial}{\partial \theta} + \frac{1}{2}\cot\theta \right) + \frac{m_\varphi}{\sin\theta}\sigma^1 \right] \xi(\theta) = \kappa \xi(\theta). \tag{21}$$

In (18) and (21), $\sigma^k$ are Pauli matrices, and

$$\kappa = \mp 1, \mp 2, ... = \begin{cases} -(l+1), & j = l+1/2 \\ l, & j = l-1/2 \end{cases}. \tag{22}$$

As a result, taking into account (18) and the definition of the spinor[2] $\xi(\theta) = \begin{pmatrix} Y_-(\theta) \\ Y_+(\theta) \end{pmatrix}$,

Eq. (17) can written as a set of four equations:

---

[2] Unlike (19), hereafter in the notations $_{\mp 1/2}Y(\theta)$ the indices $j, m_\varphi$ are dropped for shortness.

$$Y_- E\,\omega(r,\theta) = Y_- \frac{m}{\sqrt{g_{KN}^{00}}}\omega(r,\theta) -$$

$$-Y_-\left[\frac{\sqrt{\Delta_{KN}}}{r_K\sqrt{g_{KN}^{00}}}\left(\frac{\partial}{\partial r}+\frac{1}{r}\right)-\frac{\kappa}{r_K\sqrt{g_{KN}^{00}}}+\frac{1}{2}\frac{\partial}{\partial r}\left(\frac{\sqrt{\Delta_{KN}}}{r_K\sqrt{g_{KN}^{00}}}\right)\right]\chi(r,\theta) +$$

$$+Y_+ \frac{1}{r_K\sqrt{g_{KN}^{00}}}\frac{\partial}{\partial\theta}\chi(r,\theta)+Y_+\frac{1}{2}\frac{\partial}{\partial\theta}\left(\frac{1}{r_K\sqrt{g_{KN}^{00}}}\right)\chi(r,\theta) +$$

$$+Y_+\left(\frac{1}{g_{KN}^{00}\sqrt{\Delta_{KN}}}-\frac{1}{r_K\sqrt{g_{KN}^{00}}}\right)\frac{m_\varphi}{\sin\theta}\chi(r,\theta)+Y_-\frac{ar_0 r m_\varphi}{g_{KN}^{00}r_K^2\Delta_{KN}}\omega(r,\theta) +$$

$$+Y_+\frac{1}{4}\frac{\sqrt{g_{KN}^{00}}\Delta_{KN}}{r_K}ar_0\sin\theta\frac{\partial}{\partial r}\left(\frac{r}{g_{KN}^{00}r_K^2\Delta_{KN}}\right)\omega(r,\theta) -$$

$$-Y_-\frac{1}{4}\frac{\sqrt{g_{KN}^{00}}\sqrt{\Delta_{KN}}}{r_K}ar_0\sin\theta\frac{\partial}{\partial\theta}\left(\frac{r}{g_{KN}^{00}r_K^2\Delta_{KN}}\right)\omega(r,\theta)+Y_-\frac{qQ}{r}\omega(r,\theta); \qquad (23)$$

$$Y_+ E\,\omega(r,\theta) = Y_+ \frac{m}{\sqrt{g_{KN}^{00}}}\omega(r,\theta) -$$

$$-Y_+\left[\frac{\sqrt{\Delta_{KN}}}{r_K\sqrt{g_{KN}^{00}}}\left(\frac{\partial}{\partial r}+\frac{1}{r}\right)-\frac{\kappa}{r_K\sqrt{g_{KN}^{00}}}+\frac{1}{2}\frac{\partial}{\partial r}\left(\frac{\sqrt{\Delta_{KN}}}{r_K\sqrt{g_{KN}^{00}}}\right)\right]\chi(r,\theta) -$$

$$-Y_-\frac{1}{r_K\sqrt{g_{KN}^{00}}}\frac{\partial}{\partial\theta}\chi(r,\theta)-Y_-\frac{1}{2}\frac{\partial}{\partial\theta}\left(\frac{1}{r_K\sqrt{g_{KN}^{00}}}\right)\chi(r,\theta) +$$

$$+Y_-\left(\frac{1}{g_{KN}^{00}\sqrt{\Delta_{KN}}}-\frac{1}{r_K\sqrt{g_{KN}^{00}}}\right)\frac{m_\varphi}{\sin\theta}\chi(r,\theta)+Y_+\frac{ar_0 r m_\varphi}{g_{KN}^{00}r_K^2\Delta_{KN}}\omega(r,\theta) +$$

$$+Y_-\frac{\sqrt{g_{KN}^{00}}\Delta_{KN}}{4r_K}ar_0\sin\theta\frac{\partial}{\partial r}\left(\frac{r}{g_{KN}^{00}r_K^2\Delta_{KN}}\right)\omega(r,\theta) +$$

$$+Y_+\frac{\sqrt{g_{KN}^{00}}\sqrt{\Delta_{KN}}}{4r_K}ar_0\sin\theta\frac{\partial}{\partial\theta}\left(\frac{r}{g_{KN}^{00}r_K^2\Delta_{KN}}\right)\omega(r,\theta)+Y_+\frac{qQ}{r}\omega(r,\theta); \qquad (24)$$

$$Y_- E \chi(r,\theta) = -Y_- \frac{m}{\sqrt{g_{KN}^{00}}} \chi(r,\theta) +$$

$$+Y_- \left[ \frac{\sqrt{\Delta_{KN}}}{r_K \sqrt{g_{KN}^{00}}} \left( \frac{\partial}{\partial r} + \frac{1}{r} \right) + \frac{\kappa}{r_K \sqrt{g_{KN}^{00}}} + \frac{1}{2} \frac{\partial}{\partial r} \left( \frac{\sqrt{\Delta_{KN}}}{r_K \sqrt{g_{KN}^{00}}} \right) \right] \omega(r,\theta) +$$

$$+Y_+ \frac{1}{r_K \sqrt{g_{KN}^{00}}} \frac{\partial}{\partial \theta} \omega(r,\theta) + \frac{Y_+}{2} \frac{\partial}{\partial \theta} \left( \frac{1}{r_K \sqrt{g_{KN}^{00}}} \right) \omega(r,\theta) +$$

$$+Y_+ \left( \frac{1}{g_{KN}^{00} \sqrt{\Delta_{KN}}} - \frac{1}{r_K \sqrt{g_{KN}^{00}}} \right) \frac{m_\varphi}{\sin \theta} \omega(r,\theta) + Y_- \frac{ar_0 r m_\varphi}{g_{KN}^{00} r_K^2 \Delta_{KN}} \chi(r,\theta) - \quad (25)$$

$$-Y_+ \frac{\sqrt{g_{KN}^{00}} \Delta_{KN}}{4 r_K} ar_0 \sin \theta \frac{\partial}{\partial r} \left( \frac{r}{g_{KN}^{00} r_K^2 \Delta_{KN}} \right) \chi(r,\theta) -$$

$$-Y_- \frac{\sqrt{g_{KN}^{00}} \sqrt{\Delta_{KN}}}{4 r_K} ar_0 \sin \theta \frac{\partial}{\partial \theta} \left( \frac{r}{g_{KN}^{00} r_K^2 \Delta_{KN}} \right) \chi(r,\theta) + Y_- \frac{qQ}{r} \chi(r,\theta);$$

$$Y_+ E \chi(r,\theta) = -Y_+ \frac{m}{\sqrt{g_{KN}^{00}}} \chi(r,\theta) +$$

$$+Y_+ \left[ \frac{\sqrt{\Delta_{KN}}}{r_K \sqrt{g_{KN}^{00}}} \left( \frac{\partial}{\partial r} + \frac{1}{r} \right) + \frac{\kappa}{r_K \sqrt{g_{KN}^{00}}} + \frac{1}{2} \frac{\partial}{\partial r} \left( \frac{\sqrt{\Delta_{KN}}}{r_K \sqrt{g_{KN}^{00}}} \right) \right] \omega(r,\theta) -$$

$$-Y_- \frac{1}{r_K \sqrt{g_{KN}^{00}}} \frac{\partial}{\partial \theta} \omega(r,\theta) - \frac{Y_-}{2} \frac{\partial}{\partial \theta} \left( \frac{1}{r_K \sqrt{g_{KN}^{00}}} \right) \omega(r,\theta) +$$

$$+Y_- \left( \frac{1}{g_{KN}^{00} \sqrt{\Delta_{KN}}} - \frac{1}{r_K \sqrt{g_{KN}^{00}}} \right) \frac{m_\varphi}{\sin \theta} \omega(r,\theta) + Y_+ \frac{ar_0 r m_\varphi}{g_{KN}^{00} r_K^2 \Delta_{KN}} \chi(r,\theta) - \quad (26)$$

$$-Y_- \frac{\sqrt{g_{KN}^{00}} \Delta_{KN}}{4 r_K} ar_0 \sin \theta \frac{\partial}{\partial r} \left( \frac{r}{g_{KN}^{00} r_K^2 \Delta_{KN}} \right) \chi(r,\theta) +$$

$$+Y_+ \frac{\sqrt{g_{KN}^{00}} \sqrt{\Delta_{KN}}}{4 r_K} ar_0 \sin \theta \frac{\partial}{\partial \theta} \left( \frac{r}{g_{KN}^{00} r_K^2 \Delta_{KN}} \right) \chi(r,\theta) + Y_+ \frac{qQ}{r} \chi(r,\theta).$$

Then, in (23) - (26), we can get rid of derivatives $\frac{\partial}{\partial \theta} \omega(r,\theta)$ and $\frac{\partial}{\partial \theta} \chi(r,\theta)$. To this end, we multiply Eq. (23) by $Y_-(\theta)$, Eq. (24) - by $Y_+(\theta)$ and add them. We do the same with Eqs. (25) and (26). As a result, we obtain:

$$E\omega(r,\theta) = \frac{m}{\sqrt{g_{KN}^{00}}}\omega(r,\theta) - \left[\frac{\sqrt{\Delta_{KN}}}{r_K\sqrt{g_{KN}^{00}}}\left(\frac{\partial}{\partial r} + \frac{1}{r}\right) - \frac{\kappa}{r_K\sqrt{g_{KN}^{00}}} + \frac{1}{2}\frac{\partial}{\partial r}\left(\frac{\sqrt{\Delta_{KN}}}{r_K\sqrt{g_{KN}^{00}}}\right)\right]\chi(r,\theta) +$$

$$+ \left(\frac{1}{g_{KN}^{00}\sqrt{\Delta_{KN}}} - \frac{1}{r_K\sqrt{g_{KN}^{00}}}\right)\frac{m_\varphi}{\sin\theta}\frac{2Y_-Y_+}{(Y_-)^2 + (Y_+)^2}\chi(r,\theta) +$$

$$+ \frac{ar_0 rm_\varphi}{g_{KN}^{00} r_K^2 \Delta_{KN}}\omega(r,\theta) + \frac{1}{4}\frac{\sqrt{g_{KN}^{00}}\Delta_{KN}}{r_K}ar_0\sin\theta\frac{\partial}{\partial r}\left(\frac{r}{g_{KN}^{00} r_K^2 \Delta_{KN}}\right)\frac{2Y_-Y_+}{(Y_-)^2 + (Y_+)^2}\omega(r,\theta) -$$

$$- \frac{1}{4}\frac{\sqrt{g_{KN}^{00}}\sqrt{\Delta_{KN}}}{r_K}ar_0\sin\theta\frac{\partial}{\partial\theta}\left(\frac{r}{g_{KN}^{00} r_K^2 \Delta_{KN}}\right)\frac{(Y_-)^2 - (Y_+)^2}{(Y_-)^2 + (Y_+)^2}\omega(r,\theta) + \frac{qQ}{r}\omega(r,\theta);$$

(27)

$$E\chi(r,\theta) = -\frac{m}{\sqrt{g_{KN}^{00}}}\chi(r,\theta) + \left[\frac{\sqrt{\Delta_{KN}}}{r_K\sqrt{g_{KN}^{00}}}\left(\frac{\partial}{\partial r} + \frac{1}{r}\right) + \frac{\kappa}{r_K\sqrt{g_{KN}^{00}}} + \frac{1}{2}\frac{\partial}{\partial r}\left(\frac{\sqrt{\Delta_{KN}}}{r_K\sqrt{g_{KN}^{00}}}\right)\right]\omega(r,\theta) +$$

$$+ \left(\frac{1}{g_{KN}^{00}\sqrt{\Delta_{KN}}} - \frac{1}{r_K\sqrt{g_{KN}^{00}}}\right)\frac{m_\varphi}{\sin\theta}\frac{2Y_-Y_+}{(Y_-)^2 + (Y_+)^2}\omega(r,\theta) +$$

$$+ \frac{ar_0 rm_\varphi}{g_{KN}^{00} r_K^2 \Delta_{KN}}\chi(r,\theta) - \frac{1}{4}\frac{\sqrt{g_{KN}^{00}}\Delta_{KN}}{r_K}ar_0\sin\theta\frac{\partial}{\partial r}\left(\frac{r}{g_{KN}^{00} r_K^2 \Delta_{KN}}\right)\frac{2Y_-Y_+}{(Y_-)^2 + (Y_+)^2}\chi(r,\theta) -$$

$$- \frac{1}{4}\frac{\sqrt{g_{KN}^{00}}\sqrt{\Delta_{KN}}}{r_K}ar_0\sin\theta\frac{\partial}{\partial\theta}\left(\frac{r}{g_{KN}^{00} r_K^2 \Delta_{KN}}\right)\frac{(Y_-)^2 - (Y_+)^2}{(Y_-)^2 + (Y_+)^2}\chi(r,\theta) + \frac{qQ}{r}\chi(r,\theta).$$

(28)

Equations (27), (28) can be used for the standard procedure of obtaining effective potentials. In this case, the angle $\theta$ and the particle energy $E$ are parameters.

Below, we will write our expressions in dimensionless variables

$$\rho = \frac{r}{l_c}, \quad \varepsilon = \frac{E}{m}, \quad \alpha = \frac{r_0}{2l_c} = \frac{GMm}{\hbar c} = \frac{Mm}{M_P^2},$$

$$\alpha_Q = \frac{r_Q}{l_c} = \frac{\sqrt{G}Qm}{\hbar c} = \frac{\sqrt{\alpha_{fs}}}{M_P}m\frac{Q}{e}, \quad \alpha_a = \frac{a}{l_c}, \quad \alpha_{em} = \frac{qQ}{\hbar c} = \alpha_{fs}\frac{Q}{e}\frac{q}{e}.$$

(29)

Here, $l_c = \hbar/mc$ is the Compton wavelength of the Dirac particle; $M_P = \sqrt{\hbar c/G} = 2.2 \cdot 10^{-5}$ g ($1.2 \cdot 10^{19} GeV$) is the Planck mass; $\alpha_{fs} = e^2/(\hbar c) \simeq 1/137$ is the electromagnetic fine structure constant; $\alpha, \alpha_{em}$ are gravitational and electromagnetic coupling constants; $\alpha_Q, \alpha_a$ are dimensionless constants characterizing the source of the electromagnetic field and the angular momentum in the Kerr-Newman metric, respectively.

The qualities $\rho_K^2, \Delta_{KN}, g_{KN}^{00}$ in dimensionless variables have the form of

$$\rho_K^2 = \rho^2 + \alpha_a^2\cos^2\theta,$$

(30)

$$\Delta_{KN} = \rho^2 - 2\alpha\rho + \alpha_a^2 + \alpha_Q^2, \tag{31}$$

$$g_{KN}^{00} = \frac{1}{\Delta_{KN}}\left(\rho^2 + \alpha_a^2 + \frac{\alpha_a^2\left(2\alpha\rho - \alpha_Q^2\right)}{\rho_K^2}\sin^2\theta\right). \tag{32}$$

Let us introduce the following notations:

$$b = g_{KN}^{00}\Delta_{KN}\rho_K^2 = \left(\rho^2 + \alpha_a^2\cos^2\theta\right)\left(\rho^2 + \alpha_a^2\right) + \alpha_a^2\left(2\alpha\rho - \alpha_Q^2\right)\sin^2\theta, \tag{33}$$

$$\frac{\partial b}{\partial \rho} = 2\rho\left(2\rho^2 + \alpha_a^2 + \alpha_a^2\cos^2\theta\right) + \alpha_a^2 2\alpha\sin^2\theta, \tag{34}$$

$$\frac{\partial b}{\partial \theta} = -\alpha_a^2\left[\rho^2 + \alpha_a^2 - 2\alpha\rho + \alpha_Q^2\right]\sin(2\theta). \tag{35}$$

$$F(\theta) = \frac{2Y_- Y_+}{(Y_-)^2 + (Y_+)^2} = \sin\theta\frac{\left(P_l^{m_\varphi+1/2}(\cos\theta)\right)^2 - (\kappa - m_\varphi + 1/2)^2\left(P_l^{m_\varphi-1/2}(\cos\theta)\right)^2}{\left(P_l^{m_\varphi+1/2}(\cos\theta)\right)^2 + (\kappa - m_\varphi + 1/2)^2\left(P_l^{m_\varphi-1/2}(\cos\theta)\right)^2} +$$

$$+2\cos\theta\frac{(\kappa - m_\varphi + 1/2)P_l^{m_\varphi-1/2}(\cos\theta)P_l^{m_\varphi+1/2}(\cos\theta)}{\left(P_l^{m_\varphi+1/2}(\cos\theta)\right)^2 + (\kappa - m_\varphi + 1/2)^2\left(P_l^{m_\varphi-1/2}(\cos\theta)\right)^2}, \tag{36}$$

$$G(\theta) = \frac{(Y_-)^2 - (Y_+)^2}{(Y_-)^2 + (Y_+)^2} = \cos\theta\frac{(\kappa - m_\varphi + 1/2)^2\left(P_l^{m_\varphi-1/2}(\cos\theta)\right)^2 - \left(P_l^{m_\varphi+1/2}(\cos\theta)\right)^2}{(\kappa - m_\varphi + 1/2)^2\left(P_l^{m_\varphi-1/2}(\cos\theta)\right)^2 + \left(P_l^{m_\varphi+1/2}(\cos\theta)\right)^2} +$$

$$+2\sin\theta\frac{(\kappa - m_\varphi + 1/2)P_l^{m_\varphi-1/2}(\cos\theta)P_l^{m_\varphi+1/2}(\cos\theta)}{(\kappa - m_\varphi + 1/2)^2\left(P_l^{m_\varphi-1/2}(\cos\theta)\right)^2 + \left(P_l^{m_\varphi+1/2}(\cos\theta)\right)^2}. \tag{37}$$

Taking into account (29) - (37), Eqs. (27) and (28) can be rewritten as

$$\frac{\partial \omega(r,\theta)}{\partial r} = A\omega(r,\theta) + B\chi(r,\theta),$$

$$\frac{\partial \chi(r,\theta)}{\partial r} = C\omega(r,\theta) + D\chi(r,\theta). \tag{38}$$

where

$$A = -\frac{1}{\rho} - \frac{\kappa}{\sqrt{\Delta_{KN}}} - \frac{1}{\Delta_{KN}}(\rho - \alpha) + \frac{1}{4b}\frac{\partial b}{\partial \rho} - \left(\frac{\rho_K^2}{\sqrt{b\Delta_{KN}}} - \frac{1}{\sqrt{\Delta_{KN}}}\right)\frac{m_\varphi}{\sin\theta}F(\theta), \tag{39}$$

$$B = \frac{\sqrt{b}}{\Delta_{KN}}\varepsilon + \frac{\rho_K}{\sqrt{\Delta_{KN}}} - \frac{\sqrt{b}}{\Delta_{KN}}\frac{\alpha_{em}}{\rho} - \frac{2\alpha\alpha_a\rho}{\sqrt{b\Delta_{KN}}}m_\varphi - \frac{1}{2\sqrt{\Delta_{KN}}\rho_K^2}\alpha\alpha_a\sin\theta\left(1 - \frac{\rho}{b}\frac{\partial b}{\partial \rho}\right)F(\theta) -$$

$$-\frac{1}{2\Delta_{KN}\rho_K^2}\frac{\alpha\alpha_a\rho}{b}\sin\theta\frac{\partial b}{\partial \theta}G(\theta), \tag{40}$$

$$C = -\frac{\sqrt{b}}{\Delta_{KN}}\varepsilon + \frac{\rho_K}{\sqrt{\Delta_{KN}}} + \frac{\sqrt{b}}{\Delta_{KN}}\frac{\alpha_{em}}{\rho} + \frac{2\alpha\alpha_a\rho}{\sqrt{b\Delta_{KN}}}m_\varphi + \frac{1}{2\sqrt{\Delta_{KN}}\rho_K^2}\alpha\alpha_a\sin\theta\left(1 - \frac{\rho}{b}\frac{\partial b}{\partial \rho}\right)F(\theta) -$$

$$-\frac{1}{2\Delta_{KN}\rho_K^2}\frac{\alpha\alpha_a\rho}{b}\sin\theta\frac{\partial b}{\partial \theta}G(\theta), \tag{41}$$

$$D = -\frac{1}{\rho} + \frac{\kappa}{\sqrt{\Delta_{KN}}} - \frac{1}{\Delta_{KN}}(\rho - \alpha) + \frac{1}{4b}\frac{\partial b}{\partial \rho} + \left(\frac{\rho_K^2}{\sqrt{b\Delta_{KN}}} - \frac{1}{\sqrt{\Delta_{KN}}}\right)\frac{m_\varphi}{\sin\theta}F(\theta). \tag{42}$$

Then, from Eqs. (38), we obtain a second-order equation for the function $\psi_\omega(\rho,\theta)$ proportional to $\omega(\rho,\theta)$ or an equation for the function $\psi_\chi(\rho,\theta)$ proportional to $\chi(\rho,\theta)$.

$$\psi_\omega(\rho,\theta) = \omega(\rho,\theta)\exp\left(\frac{1}{2}\int_{\rho_{\min}}^{\rho} A_\omega(\rho',\theta)d\rho'\right), \tag{43}$$

$$\psi_\chi(\rho,\theta) = \chi(\rho,\theta)\exp\left(\frac{1}{2}\int_{\rho_{\min}}^{\rho} A_\chi(\rho',\theta)d\rho'\right). \tag{44}$$

In (43),

$$A_\omega(\rho,\theta) = -\frac{1}{B}\frac{\partial B}{\partial \rho} - A - D. \tag{45}$$

In (44),

$$A_\chi(\rho,\theta) = -\frac{1}{C}\frac{\partial C}{\partial \rho} - A - D. \tag{46}$$

In (43), (44), the lower limit of integration $\rho_{\min}$ is selected by the specific conditions of half-spin particle motion in the Kerr-Newman fields under consideration.

The relativistic equations for $\psi_\omega(\rho,\theta)$ and $\psi_\chi(\rho,\theta)$ have the form of the Schrödinger equation with the effective potentials $U_{eff}^\omega, U_{eff}^\chi$:

$$\frac{\partial^2 \psi_\omega}{\partial \rho^2} + 2\left(E_{Schr} - U_{eff}^\omega\right)\psi_\omega = 0, \tag{47}$$

$$\frac{\partial^2 \psi_\chi}{\partial \rho^2} + 2\left(E_{Schr} - U_{eff}^\chi\right)\psi_\chi = 0. \tag{48}$$

where

$$E_{Schr} = \frac{1}{2}(\varepsilon^2 - 1). \tag{49}$$

$$U_{eff}^\omega(\rho,\theta) = \frac{3}{8}\frac{1}{B^2}\left(\frac{\partial B}{\partial \rho}\right)^2 - \frac{1}{4}\frac{1}{B}\frac{\partial^2 B}{\partial \rho^2} + \frac{1}{4}\frac{\partial}{\partial \rho}(A-D) - \\ -\frac{1}{4}\frac{(A-D)}{B}\frac{\partial B}{\partial \rho} + \frac{1}{8}(A-D)^2 + \frac{1}{2}BC. \tag{50}$$

$$U_{eff}^\chi(\rho,\theta) = \frac{3}{8}\frac{1}{C^2}\left(\frac{\partial C}{\partial \rho}\right)^2 - \frac{1}{4}\frac{1}{C}\frac{\partial^2 C}{\partial \rho^2} - \frac{1}{4}\frac{\partial}{\partial \rho}(A-D) + \\ +\frac{1}{4}\frac{(A-D)}{C}\frac{\partial C}{\partial \rho} + \frac{1}{8}(A-D)^2 + \frac{1}{2}BC. \tag{51}$$

Equations (47), (48) and effective potentials (50), (51) are transformed into each other at $\varepsilon \to -\varepsilon, \kappa \to -\kappa, e \to -e$ $(\alpha_{em} \to -\alpha_{em})$. It follows therefrom that Eqs. (47), (48) describe the motion of Dirac particles and antiparticles. In this paper, for particles Eq. (47) is used for the function $\psi_\omega(\rho,\theta)$ with the effective potential $U_{eff}^\omega(\rho,\theta)$ (50). The nonrelativistic limit of the Dirac equation with the lower spinor, proportional to $\chi(\rho,\theta)$, disappearing at zero momentum of the particle $(\mathbf{p}=0)$, can be a basis for that. Similarly, the lower spinor with the function $\chi(\rho,\theta)$ disappears for the particle at the Foldy-Wouthuysen transformation with any value of the momentum $\mathbf{p}$ [46]. On the contrary, for an antiparticle in the nonrelativistic limit $(\mathbf{p}=0)$, at the Foldy-Wouthuysen transformation, the upper spinor of the bispinor wave function proportional to $\omega(\rho,\theta)$ disappears.

Let us note again that the polar angle $\theta$ in the expressions (39) - (48), (50), (51) is a parameter varying in the range $[0,\pi]$. The particle energy $\varepsilon$ in the potentials (50), (51) is also a parameter.

The effective potentials $U_{eff}(\rho,\theta,\alpha,\alpha_Q,\alpha_{em},\kappa,l,j,m_\varphi,\varepsilon)$, determined by expressions (50), (51), (39) - (42), have a cumbersome analytical form and can be calculated, for instance, by using the Maple software package. However, the basic features of the potential $U_{eff}$ can be analyzed manually using equations (39) - (51).

Since the potentials $U_{eff}$ parametrically depend on the angle $\theta$, unambiguous conclusions of the nature of the quantum-mechanical motion of half-spin particles can be drawn under condition of obtaining final results independent of the angle $\theta$. Otherwise, more accurate quantum-mechanical calculations are needed. This is the reason for the restricted usefulness of the method of effective potentials in the Schrödinger-type equation with nonseparable angular and radial variables.

## 5. Singularities of effective potentials for the Kerr-Newman field

An analysis of the expressions (39) - (42) and (50) shows that the effective potential $U_{eff}^\omega(\rho,\theta)$ can have isolated singularities with maximal power up to the second order. For the extreme Kerr-Newman field, this power can rise up to the fourth order.

**5.1** An important factor is the absence of any features in the effective potential associated with availability of the ergosphere (see (3), (4)). Similarly, the ergosphere availability does not manifest itself in the initial self-conjugate Hamiltonian (16) and radial equations (38).

So, the quantum mechanics of the Dirac and Schrödinger-type equations in the Kerr-Newman field with representation of the wave function in form of (18) does not anyway show the presence of the ergosphere with $g_{00} \leq 0$.

**5.2** When two horizons are present $\left(\alpha^2 > \alpha_a^2 + \alpha_Q^2\right)$, so that

$$\Delta_{KN} = (\rho - \rho_+)(\rho - \rho_-),$$
$$\rho_\pm = \alpha \pm \sqrt{\alpha^2 - \alpha_a^2 - \alpha_Q^2}, \quad (52)$$

the effective potential has second-order poles while approaching the outer or inner horizons:

$$U_{eff}^\omega \big|_{\rho \to \rho_+} \simeq \frac{3}{8} \frac{1}{B^2} \left(\frac{\partial B}{\partial \rho}\right)^2 - \frac{1}{4} \frac{1}{B} \frac{\partial^2 B}{\partial \rho^2} + \frac{1}{2} BC =$$

$$= -\frac{1}{8} \frac{1}{(\rho - \rho_+)^2} \left\{ \left[ 1 + \frac{(\varepsilon - \alpha_{em}/\rho_+)^2 \rho_+^4}{(\rho_+ - \alpha)^2} \right] + \frac{1}{(\rho_+ - \alpha)^2} \left[ \left(\varepsilon - \frac{\alpha_{em}}{\rho_+}\right)^2 \left(b\big|_{\rho_+} - \rho_+^4\right) + \frac{4\alpha^2 \alpha_a^2 \rho_+^2 m_\varphi^2}{b\big|_{\rho_+}} - \right.$$

$$\left. -4\alpha\alpha_a \rho_+ m_\varphi \left(\varepsilon - \frac{\alpha_{em}}{\rho_+}\right) - \frac{\alpha^2 \alpha_a^2 \rho_+^2 \sin^2\theta}{4\left(\rho_+^2 + \alpha_a^2 \cos^2\theta\right)^4 \left(b\big|_{\rho_+}\right)^2} \left(\frac{\partial b}{\partial \theta}\bigg|_{\rho_+}\right)^2 G^2[\theta] \right] \right\} + \quad (53)$$

$$+ O\left(\frac{1}{(\rho - \rho_+)^{3/2}}\right).$$

Here, $b\big|_{\rho_+}$ and $\frac{\partial b}{\partial \theta}\bigg|_{\rho_+}$ are the values in (33), (35) at $\rho = \rho_+$. The value $U_{eff}^\omega \big|_{\rho \to \rho_-}$ is obtained by substituting $\rho_+ \to \rho_-$ in (53).

When there is no rotation $(\alpha_a = 0)$, the summands in the second square bracket in (53) are equal to zero. The summands in the first square bracket coincide with the appropriate part of the potential for the Reissner-Nordström field [34], [35]. In this case, for charged Dirac particles at $\varepsilon \neq \alpha_{em}/\rho_\pm$, hold the conditions of quantum-mechanical "falling" to the outer and inner horizons.

In the presence of rotation $(\alpha_a \neq 0)$, such a conclusion cannot be drawn due to the presence of terms of different signs, including those depending on the angle $\theta$, in the second square bracket in (53). More accurate quantum mechanical calculations are needed for a final conclusion on the nature of half-spin particle motion near the horizons.

**5.3** The extreme Kerr-Newman field corresponds to $\left(\alpha^2 = \alpha_a^2 + \alpha_Q^2, \; \rho_+ = \rho_- = \alpha\right)$. In this case,

$$\Delta_{KN} = (\rho - \alpha)^2, \quad (54)$$

$$b\big|_{\rho\to\alpha} = \left(\alpha^2 + \alpha_a^2\right)^2, \tag{55}$$

$$\frac{\partial b}{\partial \theta}\bigg|_{\rho\to\alpha} = 0. \tag{56}$$

The effective potential in the vicinity of the single horizon $\rho_\pm = \alpha$ is

$$U_{eff}^\omega\big|_{\rho\to\alpha} = -\frac{1}{2(\rho-\alpha)^4} \times$$
$$\times \left[\left(\alpha^2 + \alpha_a^2\right)^2\left(\varepsilon - \frac{\alpha_{em}}{\alpha}\right)^2 - 4\alpha^2\alpha_a m_\varphi\left(\varepsilon - \frac{\alpha_{em}}{\alpha}\right) + \frac{4\alpha^4\alpha_a^2 m_\varphi^2}{\left(\alpha^2+\alpha_a^2\right)^2}\right] + O\left(\frac{1}{(\rho-\alpha)^3}\right). \tag{57}$$

Due to the fourth-order pole, the motion of Dirac particles in the Kerr-Newman extreme field is implemented in the mode of a "fall" to the event horizon $\rho = \alpha$.

It thus follows that in the Kerr-Newman extreme field there are no stationary bound states of half-spin particles.

For the extreme Kerr field $\left(\alpha^2 = \alpha_a^2, \alpha_Q = \alpha_{em} = 0, \rho_+ = \rho_- = \alpha\right)$, the expression (57) can be rewritten as

$$U_{eff}^\omega\big|_{\rho\to\alpha} = -\frac{2\alpha_a^4}{(\rho-\alpha)^4}\left(\varepsilon - \frac{m_\varphi}{2\alpha}\right)^2. \tag{58}$$

For the solution of

$$\varepsilon_K^{extr} = m_\varphi/2\alpha \tag{59}$$

the singularity $\sim 1/(\rho-\alpha)^4$ in the effective potential (57) disappears. However, the coefficient at the next leading singularity $\sim 1/(\rho-\alpha)^3$ depends on the angle $\theta$, and it does not allow drawing any definite conclusions. Earlier in [50], it was proved that the solution (59) is the solution for stationary bound states of half-spin particles in the extreme Kerr field.

**5.4** If there are values of the parameters $\alpha, \alpha_Q, \alpha_a, \alpha_{em}, \kappa, l, j, m_\varphi, \varepsilon, \theta$ at which expression (40) for $\rho = \rho_{cl}$ is zero

$$B(\rho = \rho_{cl}) = 0, \tag{60}$$

then, due to the first term in the effective potential (50), emerges a second-order pole with the coefficient $K = 3/8$. This pole ensures a potential barrier totally impenetrable to quantum-mechanical particles at $\rho = \rho_{cl}$ [47]. Earlier, similar barriers to Dirac particles were discovered in the repulsive Coulomb field [33] and the Reissner-Nordström field at certain values of physical parameters [35], [36].

In the general case of the Kerr-Newman field, it is extremely difficult to determine the value (or values) of $\rho_{cl}$ from Eq. (60) because of the necessity to solve an equation with polynomials of high degree. Besides, the last two summands in (40) depend on the angle $\theta$.

**5.5** For charged particles in the Kerr-Newman field, in the vicinity of the origin of coordinates $\rho = 0$, the effective potential (50) depends on $\theta$. For uncharged particles in the Kerr-Newman field $(\alpha_{em} = 0)$ and for any particles in the Kerr field $(\alpha_Q = 0, \alpha_{em} = 0)$

$$U_{eff}^{\omega}\Big|_{\rho \to 0} = \text{const.} \qquad (61)$$

Such a potential behaviour for the Kerr metric as $\rho \to 0$ assumes the existence of stationary bound states of half-spin particles.

## 6. Conclusions

In this paper, a self-conjugate Dirac Hamiltonian with a flat scalar product of wave functions was obtained for the Kerr-Newman field. Since this Hamiltonian does not permit separation of angular and radial variables, similarly to [37], we generalized the method of obtaining effective potentials while squaring Dirac equations. As a result, the polar angle $\theta$ is a parameter in the radial Schrödinger-type equation.

The resulting effective potentials have isolated singularities on the horizons and at certain parameters of the Kerr-Newman field and Dirac particles at some intermediate points along the radius.

An important observation is the absence of any features in the self-conjugate Hamiltonian (16), the radial equations (38) and the effective potential (50), associated with presence of the ergosphere (see (3), (4)), where $g_{00} \leq 0$. So, quantum mechanics of the Dirac and Schrödinger-type equations in the external Kerr-Newman field with the wave function representation in the form (18) does not, in any way, mark the presence of the ergosphere.

The basic results of our analysis of the effective potentials (50) are as follows:

**6.1** When there are horizons $(\alpha^2 > \alpha_a^2 + \alpha_Q^2)$, the effective potential has second-order poles (53) at $\rho = \rho_+$ and at $\rho = \rho_-$.

**6.2** For the Kerr-Newman extreme field $(\alpha^2 = \alpha_a^2 + \alpha_Q^2, \ \rho_+ = \rho_- = \alpha)$, the effective potential near the single horizon has a fourth-order pole (57). In this case, Dirac particles move in the mode of "falling" to the event horizon $\rho = \alpha$, which implies the impossibility of existence of stationary bound states of half-spin particles.

**6.3** For charged particles in the Kerr-Newman field the effective potential (50)

depends on $\theta$. For uncharged particles in the Kerr-Newman field $(\alpha_{em} = 0)$ and for any particles in the Kerr field $(\alpha_Q = 0, \alpha_{em} = 0)$ we have $U_{eff}^{\omega}\big|_{\rho \to 0} = $ const. Such an asymptotics behavior assumes the existence of stationary bound states of half-spin particles.

## Acknowledgements

The authors express their gratitude to A.L.Novoselova for technical assistance in preparation of the paper.